\documentclass[aps,nofootinbib,prd,eqsecnum,showpacs,showkeys,preprintnumbers,altaffilletter]{revtex4-1}

\usepackage[caption=false]{subfig}
\usepackage{graphicx}
\usepackage{amsmath}
\usepackage{amsfonts}
\usepackage{amssymb}
\usepackage{color}
\usepackage{bm}
\usepackage{mathrsfs}
\usepackage{epstopdf}
\usepackage{url}
\usepackage{footnote}
\usepackage{textcomp} 
\usepackage{dsfont}
\usepackage{ulem}

\makeatletter
\newcommand*{\rom}[1]{\expandafter\@slowromancap\romannumeral #1@}\makeatother

\newcommand{\be}{\begin{equation}}
  \newcommand{\ee}{\end{equation}}
\newcommand{\ben}{\begin{eqnarray*}}
  \newcommand{\een}{\end{eqnarray*}}
\newcommand{\bea}{\begin{eqnarray}}
  \newcommand{\eea}{\end{eqnarray}}
\newcommand{\bdm}{\begin{displaymath}}
  \newcommand{\edm}{\end{displaymath}}
\newcommand{\ba}{\begin{align}}
  \newcommand{\ea}{\end{align}}

\begin{document}

%\title{Quantisation of the holographic Ricci dark energy model within a Born Oppenheimer approximation}

\title{Quantisation of the holographic Ricci dark energy model}

\author{Imanol Albarran $^{1,2}$}
\email{imanol@ubi.pt}
\author{Mariam Bouhmadi-L\'{o}pez $^{1,2,3,4}$}
\email{{\mbox{mbl@ubi.pt. On leave of absence from UPV and IKERBASQUE.}}}
\date{\today}

\affiliation{
${}^1$Departamento de F\'{i}sica, Universidade da Beira Interior, 6200 Covilh\~a, Portugal\\
${}^2$Centro de Matem\'atica e Aplica\c{c}\~oes da Universidade da Beira Interior (CMA-UBI), 6200 Covilh\~a, Portugal\\
${}^3$Department of Theoretical Physics, University of the Basque Country UPV/EHU, P.O. Box 644, 48080 Bilbao, Spain\\
${}^4$IKERBASQUE, Basque Foundation for Science, 48011, Bilbao, Spain\\
}

\begin{abstract}

While general relativity is an extremely robust theory to describe the gravitational interaction in our Universe, it is expected to fail close to singularities like the cosmological ones. On the other hand, it is well known that some dark energy models might induce future singularities; this can be the case for example within the setup of the Holographic Ricci Dark Energy model (HRDE). On this work, we perform a cosmological quantisation of the HRDE model and obtain under which conditions a cosmic doomsday can be avoided within the quantum realm. We show as well that this quantum model not only avoid future singularities but also the past Big Bang.

\end{abstract}

\maketitle

%%%%%%%%%%%%%%%%%%%%%%%%%%%%%%%%%%%%%%%%%%%%%%%%%%%%%%%%%%%%%%%%%%

\section{Introduction}\label{Intro}

The late-time observed acceleration of the Universe has promoted several scenarios that try to describe this recent speed up of the Universe. In fact, it is well known that a Universe filled only with matter and radiation (for open, flat or closed spatial geometries) cannot expand with positive acceleration. Therefore, it is necessary to find other mechanisms/matters that could explain this feature. 
The simplest phenomenological approach consists in invoking another kind of energy density responsible for the current acceleration of the Universe which is usually dubbed dark energy (DE) being the cosmological constant the simplest option \cite{ls}.   

The common way to describe DE is via its equation of state (EoS) parameter, which is usually denoted by $\omega_{\textrm{DE}}$.  This cosmological parameter is the ratio between the pressure and the energy density of DE. It can be constant or time-dependent. By definition, DE EoS must fulfil $\omega_{\textrm{DE}}<-1/3$ at late-time to be able to describe the current speed up of the Universe. In fact, observations suggest $\omega_{\textrm{DE}} \thickapprox -1$ at present (see for example \cite{Planck:2015xua}).

Within the above framework, the  $\Lambda$CDM model is the best fit to the observational data \cite{Planck:2015xua}. This model assumes besides ordinary matter, the presence of non-baryonic matter corresponding to dark matter (DM) and a cosmological constant $\Lambda$. In fact, the EoS of  $\Lambda$ is just $\omega_{\Lambda}=-1$ and remains unchanged in time \cite{Peebles:2002gy}. In this Universe, the expansion is accelerated at late-time reaching an infinite scale factor in an infinite cosmic time where the Universe geometry is described by a de Sitter space-time.  
Planck latest results when combined with other cosmological measurements provides $ \omega_{\textrm{DE}}\thickapprox-1.006\pm0.045$ for a DE constant equation of state \cite{Planck:2015xua}. Even though this value is very close to minus one, a slight deviation from it is crucial for the asymptotic future behaviour of the Universe. In fact, if $\omega_{\Lambda}$ is slightly larger than -1, the Universe will continue expanding eternally, but if this value is slightly smaller than -1, and no matter how tiny is such a deviation, the Universe could end up in a doomsday. For example, it could happen that the energy density becomes infinite at a finite cosmic time \cite{Starobinsky:1999yw,Caldwell:1999ew,Caldwell:2003vq,Chimento:2003qy,Nojiri:2004pf,Nojiri:2005sr,Nojiri:2005sx,
BouhmadiLopez:2007qb,BouhmadiLopez:2006fu}.

Although the $\Lambda$CDM model gives the best observational fit to explain our current Universe, there are other ways that have gained great attention \cite{ls}. One of them is the holographic dark energy  scenario \cite{Li:2004rb,Fischler:1998st}  which is inspired on the holographic principle rooted in quantum gravity. We next explain briefly the ideas behind the HRDE.
 
As it is well known, the entropy of a given  closed system  with finite volume $L^3$ has an upper bound which is not proportional to its volume, but to its surface area, $L^2$ \cite{Bekenstein:1980jp,GonzalezDiaz:1983yf}. On the other hand, for an effective quantum field theory with a given ultra-violet (UV) cutoff,  $M_{\textrm{UV}}$, the entropy  of the same  system  is scaled as $L^3M^3_{\textrm{UV}}$. Therefore, there is always a scale or a length where the quantum field theory with  UV cutoff  is expected to fail. This will happen for large volumes or lengths. To overcome this problem a link between UV and infrared (IR) cutoffs  was proposed in \cite{Cohen:1998zx}:
 \begin{equation}\label{metric}
L^3M^4_{\textrm{UV}}\lesssim L M_p^2.
\end{equation}
This proposal ensures the validity of the quantum field theory within this regime. When the inequality is saturated, we can  define an energy density which is inversely proportional to the square of the characteristic length of the system. Applying these ideas to the universe give rise to the holographic dark energy scenario \cite{Li:2004rb}.

Now, the next task would be to find a suitable holographic energy density able to speed up the current Universe. This issue has been tackled in different works.
For example, taking as a characteristic length the inverse of the Hubble parameter, $l_\textrm{H}=H^{-1}$, it was found that the effective EoS for DE is equal to zero and therefore is  not a suitable proposal to describe the current Universe because it would imply an eternally decelerating Universe \cite{Li:2004rb}. Later on, the particle horizon was suggested as a characteristic length for the universe within the holographic approach $l_{\textrm{PH}}=a\int_0^t dt/a$ \cite{Hsu:2004ri}. In this case, it turns out that the effective EoS is larger than -1/3, therefore this choice  is equally unsuitable to describe our present Universe \cite{Hsu:2004ri,Li:2004rb}. Contrary to the previous proposals, the future event horizon $l_{\textrm{EH}}=a\int_t^{t_f} dt/a$ is phenomelogically viable as it fits the current observations \cite{Li:2004rb,Horvat:2004vn}, however, it has a drawback with causality; the future event horizon should not affect the current or past physical evolution of our Universe \cite{Cai:2007us}. There is a further possibility to define an holographic dark energy model which consists in taking as the square of the length characterising the Universe, the inverse of the Ricci scalar curvature \cite{Gao:2007ep}. This model was named the Holographic Ricci Dark Energy scenario (HRDE) for further generalization of the holographic DE model see \cite{Nojiri:2005pu,Granda:2008dk,Duran:2010ky,Chimento:2011pk,Chimento:2012fh,Chimento:2013se,Chimento:2013se}.

While the HRDE is suitable to describe the current acceleration of the universe as shown in \cite{Feng:2008rs,Gao:2007ep,BouhmadiLopez:2013pn,Ghaffari:2015foa,Suwa:2014pia}, it might induce a Big Rip singularity \cite{Gao:2007ep}; i.e., the scale factor, the Hubble parameter and its first cosmic time derivate blow up in a finite future cosmological time \cite{Starobinsky:1999yw,Caldwell:1999ew,Caldwell:2003vq}. This model has been observationally constrained in \cite{Xu:2010gg}. For more updated observational constraints on the HRDE we refer to Ref.~\cite{Suwa:2014pia}, where even interaction between DM and the HRDE is considered, and again,  a Big Rip is favoured observationally.

 When the Universe approach the Big Rip regime, we expect quantum effects to be important, therefore, it is necessary to do a  quantum analysis. The field of cosmological singularities has been extensively studied in quantum cosmology \cite{ck,pvm} (see \cite{Kamenshchik:2013naa} for a review on this topic). Quantum Cosmology consists in applying the quantum theory to the Universe as a whole \cite{Kiefer:2008sw}. A consistent theory of quantum gravity should in principle avoid the classical singularities prevalent in the classical theory of general relativity \cite{Dabrowski:2006dd,Kamenshchik:2007zj,BouhmadiLopez:2009pu}. Indeed, it has been shown that in some specific models, most DE related singularities can be avoided in the analogous quantum version \cite{Dabrowski:2006dd,Kamenshchik:2007zj,BouhmadiLopez:2009pu,Bouhmadi-Lopez:2013tua}.

In the present work, we address the quantisation of the HRDE model. We use the Wheeler-DeWitt (WDW) formalism for a homogenous, isotropic and spatially flat universe. The solutions to the WDW equation must obey the DeWitt boundary conditions \cite{ck,pvm}, which implies that the wave function of the Universe has to vanish close to the singularities, ensuring that the classical singularity is avoided through the quantisation procedure. We extend our analysis to the primordial Universe where a Big Bang is expected to take place from a classical point of view and where again a quantum analysis is required.

The paper is outlined as follows. In the next section, we make a brief review of the HRDE model. In section III, we present the WDW equation for a standard fluid and present the solutions for the HRDE model. Finally, in section IV we discuss the overall conclusions. In the Appendix, we include a brief explanation of the WKB method that was used to solve the Wheeler-DeWitt equation for some periods of the expansion of the Universe.

%%%%%%%%%%%%%%%%%%%%%%%%%%%%%%%%%%%%%%%%%%%%%%%%%%%%%%%%%%%%%%%%%%%

\section{The HRDE: a short review}\label{2}

We start reviewing the HRDE model. The Universe on its largest scale can be described by a FLRW Universe whose metric reads
\begin{equation}\label{metric}
ds^2=-dt^2+a^2\left(t\right)\left(\frac{dr^2}{1-kr^2}+r^2d\theta^2+r^2\sin^2\theta d\varphi\right),
\end{equation}
where $k=-1,0,1$ for open, flat and closed spatial geometries, respectively. Aside from the HRDE, our Universe will be filled with cold dark matter (CDM), baryonic matter and radiation. Therefore, the Friedmann equation reads
\begin{equation}\label{friedeq}
H^2=\frac{8\pi G}{3}\sum_i\rho_i-\frac{k}{a^2},
\end{equation}
where $H=\dot{a}/a$ is the Hubble parameter and dot denotes a derivate with respect to the cosmic time. The summation runs over radiation, matter, and HRDE. The energy densities of radiation and matter can be described as follows
\begin{equation}\label{rhodensr}
\rho_r=\frac{3 H_0^2}{8\pi G}\Omega_{r0}\left(\frac{a}{a_0}\right)^{-4},
\end{equation}
\begin{equation}\label{rhodensm}
\rho_m=\frac{3 H_0^2}{8\pi G}\Omega_{m0}\left(\frac{a}{a_0}\right)^{-3},
\end{equation}
where  $\Omega_{r0}$ and  $\Omega_{m0}$ are the dimensionless  energy density parameters at present time for radiation and the matter components (cold dark matter and baryonic matter), respectively. In turn, $H_0$ is the value of the Hubble parameter at present. For the HRDE, the expression for the energy density is the following \cite{Gao:2007ep}
\begin{equation}\label{riccidens}
\rho_R=6\tilde{\beta}\left(\dot{H}+2H^2+\frac{k}{a^2}\right),
\end{equation}
where  $\tilde{\beta}$ is a positive constant. A dot stands for derivates with respect to the cosmic time $t$. Defining a dimensionless quantity $\beta\equiv16\pi G\tilde{\beta}$, and solving the Friedmann equation, the expression for the  Ricci dark energy density is found to be \cite{Gao:2007ep}  
\begin{equation}\label{riccidens2}
\rho_R=\frac{3 H_0^2}{8\pi G}\left[\left(\frac{\beta}{2-\beta}\right)\Omega_{m0}\left(\frac{a}{a_0}\right)^{-3}+\Omega_{p0}\left(\frac{a}{a_0}\right)^{-2\left(2-\frac{1}{\beta}\right)}\right],
\end{equation}
where $\Omega_{p0}$ is an integration constant  which will quantify the effective amount of DE in the HRDE model \cite{Gao:2007ep}. Notice that the presence of radiation and spatial curvature in the Universe do not modify the previous result. The Ricci dark energy has one part that behaves as matter and another part which depends on the value of $\beta$ and plays the role of DE. The asymptotic future behaviour of the Universe depends on the values acquired by $\beta$, more precisely:
\begin{enumerate}
\item  If $1<\beta$ , the cosmic acceleration is negative. We disregard this case as it cannot describe the present Universe.
\item If  $1/2<\beta<1$, the Universe enters in an accelerating state when the HRDE dominates. The Universe is asymptotically flat in the future.
\item If  $\beta=1/2$ , the model is equivalent to the existence of a cosmological constant plus the matter contributions. We will disregard this case as it reduces to $\Lambda$CDM model.
\item If $0<\beta<1/2$, the Universe not only enters in an accelerated state, but also super accelerates ($\dot{H}>0$) in the future hitting a Big Rip; i.e. the Universe hit a singularity at a finite cosmic time.
\end{enumerate}

The observational constraints of the HRDE model favour the last case, i.e $0<\beta<1/2$ \cite{Suwa:2014pia,Xu:2010gg}, so the Universe would reach a future singularity in a finite time. In this framework, classical Einstein theory is no longer valid and it is necessary to make a quantum treatment.

%%%%%%%%%%%%%%%%%%%%%%%%%%%%%%%%%%%%%%%%%%%%%%%%%%%%%%%%

\section{Quantisation of the HRDE}\label{3}

The Wheeler-DeWitt equation is the analogous to the Schr\"{o}dinger equation taking the Universe as a whole; i.e. as the system to be analysed \cite{ck,pvm}. In order to deduce the WDW equation, the first step is to obtain the classical Hamiltonian from the usual Hilbert-Einstein action (in our case we restrict to a FLRW metric). Then the quantum operators are introduced according to the canonical quantisation procedure leading to the WDW equation \cite{ck,pvm}.
The gravitational action depends upon the chosen metric, in particular for a FLRW space-time, the Hilbert-Einstein action is given by 
\begin{equation}\label{action}
S_{HE}=\frac{1}{16 \pi G}\left[\int d^4x\sqrt{-g}\left(R-2\Lambda\right)-2\int d^3x\sqrt{-h}K\right].
\end{equation}
The second term on the right hand side of the previous equation contains the extrinsic curvature $K$ and is a compulsory boundary term to have a well-defined variational problem. The extrinsic curvature and its  trace read \cite{ck,pvm}
\begin{equation}\label{exK}
K_{ab}=\frac{1}{2N}\frac{\partial h_{ab}}{\partial t},
\end{equation}
\begin{equation}\label{traK}
K= K_{ab}h^{ab},
\end{equation}
where $N(t)$ is the lapse function and $h_{ab}$ is the induced spatial metric \cite{ck,pvm}
\begin{equation}\label{espmetric}
h_{ab}=a^2\left(t\right)\left(\frac{dr^2}{1-kr^2}+r^2d\theta^2+r^2\sin^2\theta d\varphi\right).
\end{equation}
The Lambda function introduced in the action corresponds to $\Lambda=8\pi G\rho$;  i.e. the matter Lagrangian of the Universe is described by $\Lambda(a)$. From the Hilbert-Einstein action, we obtain the following Lagrangian density (after performing an integral)
\begin{equation}\label{eq:L}
L= N\left[\frac{3\pi}{4G}\left(-\frac{a\dot{a}^2}{N^2}+k a-\Lambda(a)\frac{ a^3}{3}\right)\right].
\end{equation}
Defining 
\begin{equation}\label{eq:Ladot}
p_a\equiv\frac{\partial L}{\partial \dot{a}}=-\frac{3\pi}{2G}\left(\frac{a\dot{a}}{N}\right),
\end{equation}
we get the Hamiltonian corresponding to the action (\ref{action}) 
\begin{equation}\label{hamil}
\mathcal{H}=N\left[-\frac{G}{3\pi}\frac{p_a^2}{a}+\frac{\pi}{4 G}\Lambda\left(a\right)a^3\right],
\end{equation}
where $p_a$ is the canonical momentum corresponding to the scale factor. Notice that the canonical momentum associated with the lapse function does not appear because it constitutes a primary constraint \cite{pvm}. Therefore, the scale factor is the only variable in our cosmological problem. The curvature term is reabsorbed in the $\Lambda(a)$ function; i.e.:
\begin{equation}\label{lambda}
\Lambda\left(a\right)=3H_0^2\left[\Omega_{r0}\left(\frac{a}{a_0}\right)^{-4}+\left(\frac{2}{2-\beta}\right)\Omega_{m0}\left(\frac{a}{a_0}\right)^{-3}+\Omega_{k0}\left(\frac{a}{a_0}\right)^{-2}+\Omega_{p0}\left(\frac{a}{a_0}\right)^{-2\left(2-\frac{1}{\beta}\right)}\right].
\end{equation}
Here, $\Omega_{k0}$ is the dimensionless energy density parameter for curvature at present.

The term $p_a^2/a$ generates the operator
\begin{equation}\label{varchange}
\frac{p_a^2}{a}=-\hbar^2 \left[a^{-\frac{1}{2}}\partial_a\right]\left[a^{-\frac{1}{2}}\partial_a\right],
\end{equation}
in the quantum framework where we have choosen a factor ordering corresponding to the covariant generalization of the Laplace-Beltrami operator \cite{ck} (for alternative choices see for example \cite{BouhmadiLopez:2002qz,BouhmadiLopez:2004mp,BouhmadiLopez:2006pf}). It is useful to apply the following change of variable to remove the first order derivate from the quantum Hamiltonian operator
\begin{equation}\label{xdef}
x=\left(\frac{a}{a_0}\right)^\frac{3}{2}.
\end{equation}
Therefore, the quantum Hamiltonian operator can be written as
\begin{equation}\label{hamiloperator}
\hat{\mathcal{H}}=N\left\{\frac{3 G \hbar^2}{4\pi a_0^3}\partial_x^2+\frac{3\pi H_0^2 a_0^3}{4G}\left[\Omega_{r0}x^{-\frac{2}{3}}+\left(\frac{2}{2-\beta}\right)\Omega_{m0}+\Omega_{k0}x^{\frac{2}{3}}+\Omega_{p0}x^{-\frac{2}{3}\left(1-\frac{2}{\beta}\right)}\right]\right\}.
\end{equation}
As the variation of the Hamiltonian with respect to the lapse function $N$ produces the Hamiltonian constraint, the WDW equation reads
\begin{equation}\label{constrain}
\hat{\mathcal{H}}\Psi\left(x\right)=0.
\end{equation}
We will take the case of a spatially flat universe ($\Omega_{k0}=0$) for simplicity and in accordance with the current observations \cite{Planck:2015xua}. Therefore the WDW equation reduces to 
\begin{equation}\label{difeq}
\left\{\partial_x^2+\gamma\left[\Omega_{r0}x^{-\frac{2}{3}}+\left(\frac{2}{2-\beta}\right)\Omega_{m0}+\Omega_{p0}x^{-\frac{2}{3}\left(1-\frac{2}{\beta}\right)}\right]\right\}\Psi\left(x\right)=0,
\end{equation}
where we have introduced the following  dimensionless constant:
\begin{equation}\label{gamma}
\gamma\equiv\left({\pi H_0a_0^3/G\hbar}\right)^2.
\end{equation}
Due to the complexity of the equation (\ref{difeq}), we will divide the evolution of the Universe thereof in three regimes corresponding to the domination eras of radiation, matter and DE, respectively. The first one, has an exact solution, for the others two cases we will  make a WKB approximation (up to first order).
In the radiation dominated era the WDW is
\begin{equation}\label{rad}
\left\{\partial_x^2+\gamma\Omega_{r0}x^{-\frac{2}{3}}\right\}\Psi\left(x\right)=0,
\end{equation}
whose exact solution reads \cite{Ca}
\begin{equation}\label{solrad}
\Psi_1\left(x\right)=\left(\Omega_{r0}\gamma\right)^\frac{3}{8}\sqrt{x}\left[C_1J_{\frac{3}{4}}\left(\frac{3}{2}\sqrt{\Omega_{r0}\gamma}x^{\frac{2}{3}}\right)+C_2Y_{\frac{3}{4}}\left(\frac{3}{2}\sqrt{\Omega_{r0}\gamma}x^{\frac{2}{3}}\right)\right],
\end{equation}
where $C_1$ and $C_2$ are constants. The functions  $J_{3/4}$ and $Y_{3/4}$ correspond to the first and second kind Bessel functions of order $3/4$, respectively. We  choose $C_2=0$ to ensure that the wave function vanishes when $a\rightarrow0$, according with the DeWitt boundary condition \cite{Xu:2010gg,ck}.
For the matter dominated era the WDW is
\begin{equation}\label{radmat}
\left\{\partial_x^2+\gamma g_{2}\left(x\right)\right\}\Psi\left(x\right)=0,
\end{equation}
where the function $g_2 (x)$ is defined as
\begin{equation}\label{gIIxradmat}
g_{2}\left(x\right)=\left[\Omega_{r0}x^{-\frac{2}{3}}+\left(\frac{2}{2-\beta}\right)\Omega_{m0}\right].
\end{equation}
The first order WKB solution gives 
\begin{equation}\label{solradmat}
\Psi_2\left(x\right)\thickapprox\left[-\gamma g_{2}\left(x\right)\right]^{-\frac{1}{4}}\left[\alpha_1e^{ih_{2}\left(x\right)}+\alpha_2e^{-ih_{2}\left(x\right)}\right],
\end{equation}
where $\alpha_1$ and $\alpha_2$ are constants and the function $h_2 (x)$ is given by
\begin{equation}\label{hIIxradmat}
h_{2}\left(x\right)=\sqrt{\gamma\left(\frac{2}{2-\beta}\right)\Omega_{m0}}\left[x^{\frac{2}{3}}+\left(\frac{2-\beta}{2}\right)\frac{\Omega_{r0}}{\Omega_{m0}}\right]^{\frac{3}{2}}.
\end{equation}
Finally, during the DE dominated era the WDW is;
\begin{equation}\label{matde}
\left\{\partial_x^2+\gamma g_{3}\left(x\right)\right\}\Psi\left(x\right)=0,
\end{equation}
where the function $g_3(x)$ is defined as
\begin{equation}\label{gIIImatde}
g_{3}\left(x\right)=\left[\left(\frac{2}{2-\beta}\right)\Omega_{m0}+\Omega_{p0}x^s\right].
\end{equation}
The first order WKB approximation gives the solution
\begin{equation}\label{solmatde}
\Psi_3\left(x\right)\thickapprox\left[-\gamma g_{3}\left(x\right)\right]^{-\frac{1}{4}}\left[\delta_1e^{ih_{3}\left(x\right)}+\delta_2e^{-ih_{3}\left(x\right)}\right],
\end{equation}
where $\delta_1$ and $\delta_2$ are constants. The function $h_3(x)$ reads
\begin{equation}\label{hIIIxmatde}
h_{3}\left(x\right)=\frac{\sqrt{\gamma}}{2+s}x\left\{2\sqrt{g_{3}\left(x\right)}+s\ \sqrt{\left(\frac{2}{2-\beta}\right)\Omega_{m0}} \ \ {\scriptstyle2}F{\scriptstyle1}\left[\frac{1}{2},\frac{1}{s};1+\frac{1}{s};\left(\frac{\beta-2}{2}\right)\frac{\Omega_{p0}}{\Omega_{m0}}x^s\right]\right\},
\end{equation}
where $s\equiv-2/3(1-2/\beta)$.
Notice that (i) the hypergeometric function defined in (\ref{hIIIxmatde}) is not well defined as a series but it is well defined as an integral \cite{Ca} and (ii) the parameter $s$ is positive, in fact it is larger than 2 for $0<\beta<1/2$.
The function $h_3(x)$ is always real, on the other hand, the function $g_3(x)$ is positive and an increasing function of $x$; in fact, it blows up for large values of $x$. Therefore, the wave-function (\ref{solmatde}) which corresponds to an Universe dominated by DE vanishes at large scales and the DeWitt condition is fullfilled authomatically for $\Psi_3$.

Now, it is necessary to connect continuously the solutions among them. Taking the arbitrary constant $C_1=1$ (for convenience, this election will not modify the fundamental behaviour of the wave function), the conditions for a smooth wave-function give the values of $\alpha_1,\alpha_2,\delta_1,\delta_2$. These conditions are just the continuity conditions of the wave function and its first derivate on a first connecting point ($x_1$) and on a second connecting point ($x_2$), which can be read as
\begin{eqnarray}\label{cond1}
\Psi_1\left(x_1\right)=\Psi_2\left(x_1\right),
{\Psi}_{1}\prime\left(x_1\right)=\Psi_2\prime\left(x_1\right),
\Psi_2\left(x_2\right)=\Psi_3\left(x_2\right),
\Psi_{2}\prime\left(x_2\right)=\Psi_3\prime\left(x_2\right),
\end{eqnarray}
where prime stands for a derivative with respect to $x$.

We choose the first connecting point ($x_1$) in which the matter component  is  subdominant with respect to the radiation component (for example $\rho_m\sim10^{-4}\rho_r$), the second connecting  point ($x_2$) is just where the radiation and the phantom contribution in equation (\ref{difeq}) are subdominant with respect to dark matter (for example when they are equal).
 The connecting points then read
 \begin{equation}\label{x1anal}
x_1=10^{-6}\left[\left(\frac{2}{2-\beta}\right)\frac{\Omega_{m0}}{\Omega_{r0}}\right]^{-\frac{3}{2}},
\end{equation}
\begin{equation}\label{x2anal}
x_2=\left(\frac{\Omega_{r0}}{\Omega_{p0}}\right)^{\frac{3\beta}{4}}.
\end{equation}
Using the Cramer method to solve the algebraic system, the constants $\alpha_1,\alpha_2,\delta_1,\delta_2$ can be written as follows
  \begin{equation}\label{alpha1anal}
\alpha_1=-\frac{1}{2}\left[-\gamma g_2\left(x_1\right)\right]^{\frac{3}{4}}\left|\begin{array}{cc}y_{1}&a_{12}\\y_{2}&a_{22}\end{array}\right|,
\end{equation}
 \begin{equation}\label{alpha2anal}
\alpha_2=-\frac{1}{2}\left[-\gamma g_2\left(x_1\right)\right]^{\frac{3}{4}}\left|\begin{array}{cc}a_{11}&y_{1}\\a_{21}&y_{2}\end{array}\right|,
\end{equation}
 \begin{equation}\label{epsilon1anal}
\delta_1=-\frac{1}{2}\left[-\gamma g_3\left(x_2\right)\right]^{\frac{3}{4}}\left|\begin{array}{cc}z_{1}&b_{12}\\z_{2}&b_{22}\end{array}\right|,
\end{equation}
 \begin{equation}\label{epsilon2anal}
\delta_2=-\frac{1}{2}\left[-\gamma g_3\left(x_2\right)\right]^{\frac{3}{4}}\left|\begin{array}{cc}b_{11}&z_{1}\\b_{21}&z_{2}\end{array}\right|,
\end{equation}
 where we define
\begin{equation}\label{a11}
a_{11}\equiv\left[-\gamma g_2\left(x_1\right)\right]^{-\frac{1}{4}}e^{ih_2\left(x_1\right)},
\end{equation}
\begin{equation}\label{a12}
a_{12}\equiv\left[-\gamma g_2\left(x_1\right)\right]^{-\frac{1}{4}}e^{-ih_2\left(x_1\right)},
\end{equation}
\begin{equation}\label{a21}
a_{21}\equiv\left\{-\frac{\gamma^{-\frac{1}{4}}}{4}\left[-g_2\left(x_1\right)\right]^{-\frac{5}{4}}\left(\frac{2}{3}\Omega_{r0}x_1^{-\frac{5}{3}}\right)-i\left[\gamma g_2\left(x_1\right)\right]^{-\frac{1}{2}}\right\}e^{ih_2\left(x_1\right)},
\end{equation}
\begin{equation}\label{a22}
a_{22}\equiv\left\{-\frac{\gamma^{-\frac{1}{4}}}{4}\left[-g_2\left(x_1\right)\right]^{-\frac{5}{4}}\left(\frac{2}{3}\Omega_{r0}x_1^{-\frac{5}{3}}\right)+i\left[\gamma g_2\left(x_1\right)\right]^{-\frac{1}{2}}\right\}e^{-ih_2\left(x_1\right)},
\end{equation}
\begin{equation}\label{y1}
y_1\equiv\left(\Omega_{r0}\gamma\right)^{\frac{3}{8}}\sqrt{x_1}J_{\frac{3}{4}}\left(\frac{3}{2}\sqrt{\Omega_{r0}\gamma}x_{1}^{\frac{2}{3}}\right),
\end{equation}
\begin{equation}\label{y2}
y_2\equiv\left(\Omega_{r0}\gamma\right)^{\frac{3}{8}}\frac{1}{2\sqrt{x_1}}\left[J_{\frac{3}{4}}\left(\frac{3}{2}\sqrt{\Omega_{r0}\gamma}x_{1}^{\frac{2}{3}}\right)-2\left(\Omega_{r0}\gamma\right)^{\frac{1}{2}}x_1^{-\frac{1}{3}}J_{\frac{7}{4}}\left(\frac{3}{2}\sqrt{\Omega_{r0}\gamma}x_{1}^{\frac{2}{3}}\right)+1\right],
\end{equation}
\begin{equation}\label{b11}
b_{11}\equiv\left[-\gamma g_3\left(x_2\right)\right]^{-\frac{1}{4}}e^{ih_3\left(x_2\right)},
\end{equation}
\begin{equation}\label{b12}
b_{12}\equiv\left[-\gamma g_3\left(x_2\right)\right]^{-\frac{1}{4}}e^{-ih_3\left(x_2\right)},
\end{equation}
\begin{equation}\label{b21}
b_{21}\equiv\left\{\frac{\gamma^{-\frac{1}{4}}}{4}\left[-g_3\left(x_2\right)\right]^{-\frac{5}{4}}\left(s\Omega_{p0}x_2^{s-1}\right)-i\left[\gamma g_3\left(x_2\right)\right]^{-\frac{1}{2}}\right\}e^{ih_3\left(x_2\right)},
\end{equation}
\begin{equation}\label{b22}
b_{22}\equiv\left\{\frac{\gamma^{-\frac{1}{4}}}{4}\left[-g_3\left(x_2\right)\right]^{-\frac{5}{4}}\left(s\Omega_{p0}x_2^{s-1}\right)+i\left[\gamma g_3\left(x_2\right)\right]^{-\frac{1}{2}}\right\}e^{-ih_3\left(x_2\right)},
\end{equation}
\begin{equation}\label{z1}
z_1\equiv\left[-\gamma g_{2}\left(x_2\right)\right]^{-\frac{1}{4}}\left[\alpha_1e^{ih_{2}\left(x_2\right)}+\alpha_2e^{-ih_{2}\left(x_2\right)}\right],
\end{equation}
\begin{eqnarray}\label{z2}
z_2\equiv\alpha_1\left\{-\frac{\gamma^{-\frac{1}{4}}}{4}\left[-g_2\left(x_2\right)\right]^{-\frac{5}{4}}\left(\frac{2}{3}\Omega_{r0}x_2^{-\frac{5}{3}}\right)-i\left[\gamma g_2\left(x_2\right)\right]^{-\frac{1}{2}}\right\}e^{ih_2\left(x_2\right)} \\
-\alpha_2\left\{\frac{\gamma^{-\frac{1}{4}}}{4}\left[-g_2\left(x_2\right)\right]^{-\frac{5}{4}}\left(\frac{2}{3}\Omega_{r0}x_2^{-\frac{5}{3}}\right)+i\left[\gamma g_2\left(x_2\right)\right]^{-\frac{1}{2}}\right\}e^{-ih_2\left(x_2\right)}.
\end{eqnarray}
With the aim to show a numerical  result, we use the constraints $\beta=0.3823$, $\Omega_{r0}=8\cdot10^{-5}, \Omega_{m0}=0.2927, \Omega_{p0}=0.6380$ in accordance with the best fit of the current Universe within the HRDE model \cite{Xu:2010gg}. Therefore, the connecting points are 
\begin{equation}\label{metp}
 x_1=3.287\cdot 10^{-12} , \  x_2=0.07608,
\end{equation}
and the constants can be written as
\begin{equation}\label{constants}
\alpha_1=u_{\alpha}+v_{\alpha}i, \\
\alpha_2=v_{\alpha}+u_{\alpha}i, \\
\delta_1=u_{\delta}+v_{\delta}i, \\
\delta_2=v_{\delta}+u_{\delta}i.
\end{equation}
We next show an example of these constants. We choose $\gamma=10^{20}$, sufficiently large to ensure the validity of the WKB approximation for the connecting first point; i.e. well inside the radiation dominated epoch where matter is subdominant. Notice that if the WKB approximation is fulfilled for the first connecting point it is well defined for the second connecting point. Then we obtain $q(x_1 )=0.035$ (see Eq.~(\ref{ineqWKB})). Finally, using the above introduced constraints and for the selected $\gamma$ we find: 
\begin{equation}\label{results}
u_{\alpha}=80274.39, \\
v_{\alpha}=-288039.58, \\
u_{\delta}=-298682.24, \\
v_{\delta}=14131.34.
\end{equation}
In principle, $\gamma$ is much larger, in fact, taking $a_0=1$ gives the value $\gamma=8.072\cdot10^{87}$.  If in addition, we take into account the whole number of e-folds since the radiation dominated epoch, the parameter $\gamma$ would be much larger. We choose the previous value of $\gamma$; i.e. $\gamma=10^{20}$, to be able to get numerically the values of the constants defined in Eq. (\ref{constants}).  

%still have to introduce the Universe size in (\ref{gamma}) to achieve the real value.

%$C=0.2846$

%H_0^2=69.08 (\textrm{Km/Mpc})%

%\begin{equation}\label{results}
%u_{\alpha}=22846.49, \nonumber\\
%v_{\alpha}=-81977.50, \nonumber\\
%u_{\delta}=-85006.46, \nonumber\\
%v_{\delta}=4021.85.
%\end{equation}

%%%%%%%%%%%%%%%%%%%%%%%%%%%%%%%%%%%%%%%%%%%%%%%%%%%%%%%%%%%%%%%%%%

\section{Conclusions and outlook}
\label{end}

The HRDE is a suitable proposal to describe the late Universe. The best fit of the model provides a value for the proportionality constant $\beta$ inside the interval $0<\beta<1/2$ \cite{Xu:2010gg}. This means that the Universe is not only accelerating but will also face a future Big Rip singularity. This classical singularity is analysed within a quantum treatment where the quantisation is realised in the framework of the WDW equation for a flat FLRW universe and imposing the DeWitt boundary condition. We have shown that the Big Rip singularity as a consequence of the phantom like behaviour of the HRDE could be avoidable and would be harmless within the quantum approach used in this work. This might not be the case within a classical approach \cite{delCampo:2013hka}.  The DW condition can be regarded as a guidance in the nowadays incomplete theory of quantum cosmology. In fact, the disappearance of the probability distribution at singularities should arise in the theory in a natural way as a dynamical consequence of some other requirements, such as the normalizability of the wave function, and should not be postulated. Given that we lack of a  complete and consistent quantum gravity theory, we will stick to the DW condition as our guidance for singularity avoidance. 

Despite the quantum analysis presented in this work about the avoidance of the Big Bang and Big Rip singularities, this fact cannot be interpreted as an exact and thorough evidence of singularities avoidance in quantum cosmology, but rather an indication that a consistent and complete quantum theory of gravity should be free of these singularities.

To solve the Wheeler-DeWitt equation we carry two types of approximations: (i) we divide the evolution of the universe in three different epoch corresponding to radiation, matter and DE dominance as explained in the previous section, (ii) for the last two periods we use a WKB approximation where it is enough to go to first order in the WKB approximation to ensure the DeWitt boundary condition for large scale factors where the wave function is asymptotically decreasing and vanishing.

During the radiation dominated epoch, we obtain the exact wave function, fulfilling the DeWitt condition, which can be matched with the WKB approximation for the second period (matter+radiation). The larger is the value of $\gamma$, defined in (\ref{gamma}), the sooner, i.e. for smaller scale factors, we can carry the matching between the two wave functions. In fact, the asymptotic behaviour of the wave function during the radiation dominated epoch matches smoothly and naturally with the WKB approximated solution corresponding to the matter and radiation epochs.

The quantisation is necessary to describe the Universe close to singularities. However, outside of these singularities the Universe can be described classically. In this regime, the square of the modulus of the wave function can be interpreted as the classical probability density distribution. We can see this clearly when the value of $\gamma$ is very large and therefore, the wave function carries out a significant number of oscillations within a short interval of the chosen variable. In fact, these oscillations are modulated by the classical probability density distribution, which is defined as the time average of the scale factor.

In an analogy with the classical picture, a slight deviation of the resulted wave function from the classical probability density distribution is expected. This is due to the performed first type of approximation, which disregards the contribution of DE for small scales and the contribution of radiation for large scales. Therefore, this deviation is more significant in the matter dominated epoch.

Next, we would like to stress that our model has only one degree of freedom   described through the scale factor which in fact can play the role of the classical time. On the other hand,  gravity is a reparametrisation invariant theory with first class constraints. Therefore, there is always a gauge fixing condition which reduces the number of physical variables and in our case we would be left without any degree of freedom after the gauge fixing. However, our  current work can be regraded as a first approach in quantising the HRDE model, in fact  a toy model, and we will present in a different work a more elaborated scenario for the quantisation of the  HRDE with two physical variables and a genuine degree of freedom \cite{iam}. In fact, to get at least one degree of freedom with physical meaning, we can map the matter content given by a perfect fluid to one or more scalar fields that could mimic the different components of the universe. This method was carried out for example in \cite{Kamenshchik:2012ij} for a minimally coupled  scalar field or a tachyon scalar field. 

In addition, we have been mainly focussing on the wave function of the universe while in fact the important thing is the probabilistic interpretation of it. This requires the definition of a Hilbert space with a proper scalar product and measure. In the interesting review \cite{Barvinsky:1993jf}, this non trivial problem is discussed and several potential solutions are presented while in \cite{Kamenshchik:2013naa,Barvinsky:2013aya} those procedures are applied to different  cosmological models. We hope to implement those methods for the HRDE in a future work.

The aforementioned interpretation has some conceptual drawbacks: first, this picture corresponds with the description of an ``external observer'', who lives in a reality outside from the quantum system under study. This is certainly not the case for a cosmological system where the observer is part of the system \cite{Barvinsky:1993jf}.  In fact, we would need to apply a quantum theory of closed systems or the many-worlds interpretation of quantum mechanics to our universe but this is beyond the scope of our current work (see \cite{Barvinsky:1993jf} for more details on this subject). 

We have restricted our analysis to a homogeneous and isotropic configuration. However, it is well known that the inhomogeneities and anisotropies can be quite important close to singularities, for example through the creation of particle and gravitons (see for example \cite{Barrow:2011ub,Tavakoli:2014mra}). We leave this interesting issue for a future work.

%%%%%%%%%%%%%%%%%%%%%%%%%%%%%%%%%%%%%%%%%%%%%%%%%%%%%%%%%%%%%%%%%

\section*{Acknowledgments}

IA is supported by the Grant PTDC/FIS/111032/2009. He acknowledges the hospitality of the departament of theoretical physics and history of science of the University of the Basque Country where part of the work was carried out. He also wishes to acknowledge the careful reading of F. Cabral and S. Kumar of a previous version of the work. The work of MBL is supported by the Portuguese Agency ``Funda\c{c}\~{a}o para a Ci\^{e}ncia e Tecnologia" through an Investigador FCT Research contract, with reference IF/01442/2013/CP1196/CT0001. She also wishes to acknowledge the support from the Portuguese Grants PTDC/FIS/111032/2009 and UID/MAT/00212/2013  and the partial support from the Basque government Grant No. IT592-13 (Spain).

\appendix

\section{Justification of the WKB approximation}\label{justBO}

For the following differential equation

\begin{equation}\label{difeqWKB}
\left\{\partial_x^2+\gamma g\left(x\right)\right\}\Psi\left(x\right)=0,
\end{equation}
the first order WKB approximation gives the solution \cite{wkbref} 
\begin{equation}\label{solWKB}
\Psi\left(x\right)\thickapprox\left(g\left(x\right)\right)^{-\frac{1}{4}}\left[A_1e^{h\left(x\right)}+A_2e^{-h\left(x\right)}\right],
\end{equation}
where $A_1$ and $A_2$ are  constants of integration and $h(x)$ reads
\begin{equation}\label{hWKB}
h\left(x\right)=\int\sqrt{-\gamma g\left(x\right)}dx.
\end{equation}
The first order WKB approximation is valid in the region that complies with the following inequality \cite{wkbref}
\begin{equation}\label{ineqWKB}
q\left(x\right)\equiv\frac{1}{\gamma}\left|\frac{5\left[g\prime\left(x\right)\right]^2-4\left[g\prime\prime\left(x\right)\right]\left[g\left(x\right)\right]}{16\left[g\left(x\right)\right]^{3}}\right| \ll1.
\end{equation}
For the chosen interval (\ref{metp}), this inequality is fulfilled for large values of the dimensionless quantity $\gamma$. On the previous expression a prime stands for a derivative with respect to $x$.

%%%%%%%%%%%%%%%%%%%%%%%%%%%%%%%%%%%%%%%%%%%%%%%%%%%%%%%%%%%%%%%%%%%%%%%%%%%%%

\end{document}